\pdfoutput=1
\documentclass[useAMS,usenatbib]{mn2e}
\usepackage[pdftex]{graphicx}
\usepackage{times}
\usepackage{wasysym}
\usepackage{url}
\usepackage{lineno}
\usepackage{amssymb}

\newcommand{\dg}{DG\,CVn}
\newcommand{\g}{$\gamma$}

\title{On the expected \g-ray emission from nearby flaring stars}

\author[S. Ohm and C. Hoischen]{\normalsize S.~Ohm$^{1}$ and C.~Hoischen$^{2}$\\
  $^{1}$ DESY, D-15738 Zeuthen, Germany \\
  $^{2}$ Institut f\"ur Physik und Astronomie, Universit\"at Potsdam, Karl-Liebknecht-Strasse 24/25, D 14476 Potsdam, Germany \\
}

\begin{document}

\date{Accepted 2017 October 24. Received 2017 October 16; in original form
  2017 September 18}
\pagerange{\pageref{firstpage}--\pageref{lastpage}}
\pubyear{2017}

\maketitle

\label{firstpage}

\begin{abstract} {Stellar flares have been extensively studied in soft
    X-rays (SXR) by basically every X-ray mission. Hard X-ray (HXR)
    emission from stellar superflares, however, have only been
    detected from a handful of objects over the past years. One very
    extreme event was the superflare from the young M-dwarf \dg\
    binary star system, which triggered Swift/BAT as if it was a
    \g-ray burst (GRB). In this work, we estimate the expected \g-ray
    emission from \dg\ and the most extreme stellar flares by
    extrapolating from solar flares based on measured solar energetic
    particles (SEPs), as well as thermal and non-thermal emission
    properties. We find that ions are plausibly accelerated in stellar
    superflares to 100\, GeV energies, and possibly up to TeV energies
    in the associated coronal mass ejections. The corresponding
    $\pi^0$-decay \g-ray emission could be detectable from stellar
    superflares with ground-based \g-ray telescopes. On the other
    hand, the detection of \g-ray emission implies particle densities
    high enough that ions suffer significant losses due to inelastic
    proton-proton scattering. The next-generation Cherenkov Telescope
    Array (CTA) should be able to probe superflares from M-dwarfs in
    the solar neighbourhood and constrain the energy in interacting
    cosmic rays and/or their maximum energy. The detection of \g-ray
    emission from stellar flares would open a new window for the study
    of stellar physics, the underlying physical processes in flares
    and their impact on habitability of planetary systems.}
\end{abstract}


\maketitle

\section{Introduction}\label{sec:intro}

Flares are a common phenomenon in the Sun, in stars and young stellar
objects \citep[see e.g.][for a recent review]{Benz2010}. They
typically result from a restructuring of the stellar magnetic field
and manifest themselves in a release of magnetic energy by
reconnection. In these reconnection events, ionising radiation is
produced and particles are accelerated to non-thermal
energies. Flares, especially in young stars, are important as the
released radiation and energetic particles can influence outer
planetary atmospheres by heating and ionisation
(e.g. \citet{Segura2010, Venot2016}). The most extreme flares to date
are observed in X-rays from nearby M-dwarf stars.

\dg\ is a young (30\,Myr) M-dwarf binary star system located at 18\,pc
distance to the Sun. At least one of its member stars is a rapid
rotator. This system underwent an extreme flaring event on April 23rd,
2014 \citep{Drake2014}, which triggered the Burst Alert Telescope
(BAT) aboard the {\it Swift} satellite, and caused it to slew
automatically to the source as if it was a GRB. The BAT detected \dg\
in the hard X-ray band between $15 - 150$\,keV with a peak flux of
$\sim$300\,mCrab. The \g-ray superflare is one of only a handful
detected by {\it Swift} in a decade, and was $\sim$200 thousand times
more energetic in X-rays than X-class solar flares (class X1
corresponds to $3\times10^{26}$\,erg\,s$^{-1}$). The first flaring
episode lasted for 2.2\,hours, and had a total energy in the
$0.3 - 10$\,keV energy band of $4\times 10^{35}$\,erg. The day after,
a second flare occurred, which had a similar radiated energy as the
first one, and lasted for 6.4\,hours \citep{Osten2016}. Different
episodes of the superflare were observed by the XRT and UVOT
instrument onboard {\it Swift} after the repositioning. Also other
ground-based optical instruments followed-up on the BAT alert. The
results of these observations are reported in detail elsewhere
\citep{Osten2016, Caballero2015}. As part of the AMI-LA Rapid Response
Mode (ALARRM), prompt radio follow-up observations were taken at 15
GHz within six minutes of the burst. \cite{Fender2015} reported on the
detection of a bright ($\sim$100\,mJy) radio flare in these
observations. An additional bright radio flare, peaking at around 90
mJy, occurred around one day later. In high-energy \g\ rays
(100\,MeV\,$\leq E \leq$\,100\,GeV), \cite{Loh2017} searched for a
counterpart in archival {\it Fermi}-LAT data. The authors found a
transient event (but happening in 2012), at a position consistent with
\dg, which they associate to be a likely flaring background
blazar. The imaging atmospheric Cherenkov telescope (IACT) system
MAGIC operates in the very-high-energy domain
(50\,GeV\,$\leq E \leq$\,50\,TeV) and followed up on this flare within
30 seconds after the alert and observed it for 3.3 hours
\citep{MAGIC2014}. The authors placed an upper limit on the energy
flux above 200\,GeV of $1.2\cdot10^{-11}$\,erg\,cm$^{-2}$\,s$^{-1}$.

Given the characteristics of the superflare from \dg\ and other
systems, we argue in this letter that events like the one in 2014 are
potentially detectable by current- and next-generation IACTs like
MAGIC, VERITAS, H.E.S.S., or the future Cherenkov Telescope Array
(CTA). In section \ref{sec:solarflares} we shortly review the
properties of solar flares with focus on particle acceleration and
high-energy \g-ray emission. Section \ref{sec:stellarflares} discusses
if and how properties of solar flares can be extrapolated to stellar
flares. In section \ref{sec:model} we present two possible scenarios
for particle acceleration in the \dg\ event and apply a non-thermal
emission model to estimate the expected \g-ray emission in the tens of
GeV energy range and above. In section \ref{sec:cta} we will extend
the non-thermal emission model to the population of nearby young dwarf
stars and estimate the number of potentially detectable stellar flares
with the upcoming CTA.

\section{Solar energetic particles and \g-ray production in solar
  flares}\label{sec:solarflares}

Most of the knowledge about physical mechanisms at work in stellar
flares has been acquired in studies of solar flares. Especially highly
relativistic particles that are released in stellar flares are
notoriously difficult to study. Radio cyclotron and gyrosynchrotron
emission originates from thermal and mildly relativistic electrons up
to MeV energies. Higher energy electrons can in principle be studied
via their non-thermal X-ray emission. But the hard X-ray (HXR)
emission detected in stellar flares is better described as the tail of
a thermal hot plasma ($50-100$\,MK), rather than a non-thermal
thick-target bremsstrahlung component. At \g-ray energies, no stellar
flare has been detected to date and studies of relativistic protons
and nuclei and their impact on the surroundings of these stars are
limited to extrapolations from solar flares
\citep[e.g.][]{Segura2010}.

Solar flares are observed across the electromagnetic spectrum from
radio to \g\ rays. Contrary to stellar flares, non-thermal HXR
synchrotron emission from relativistic electrons has been detected
from the Sun up to tens of MeV. At \g-ray energies, flare-accelerated
ions excite heavier nuclei when impinging on the chromosphere, and
subsequently emit nuclear \g-ray lines. Neutrons produced in the
spallation of nuclei in large flares are captured by protons, which
produce the deuterium recombination line at 2.22\,MeV. This
recombination line allows one to infer the flux, spectrum and total
energy of ions. At even higher energies of 100\,MeV and above,
continuum radiation is emitted by ions inelastically scattering with
ambient nuclei, which produce charged and neutral pions. Neutral pions
decay into two \g\ rays which can be studied with instruments such as
the {\it Fermi}-LAT. Over the past couple of years, the \g-ray
emission at hundreds of MeV to GeV energies from solar flares has been
studied in quite some detail. The quiescent sun has been studied by
\citet{Abdo2011}, whereas solar flares have been investigated by
\citet{Ackermann2012, Ajello2014} and
\citet{Ackermann2014}. \citet{Pesce-Rollins2015} and
\citet{Ackermann2017} reported on three behind-the-limb (BTL) solar
flares. The findings of these works can be summarised as follows: i)
Particle acceleration is more common in solar flares than previously
thought. ii) The short, impulsive phase of flares can be followed by a
longer duration phase that can last up to several hours. iii) \g-ray
emission is associated with flares that are accompanied by fast
coronal mass ejections (CMEs). iv) Accelerated protons better describe
the spectral energy distribution than accelerated electrons. v)
Maximum \g-ray energies of several GeV imply the presence of tens of
GeV protons and nuclei in the acceleration region.

Two main scenarios for the acceleration of solar energetic particles
(SEPs) are established by now: Impulsive events, triggered by magnetic
reconnection, produce steep particle spectra
($dN/dE \propto E^{-\alpha}$, $\alpha \simeq 3.0$) that are highly
enriched in $^{3}$He and heavier ions. Gradual SEP events, on the
other hand, are caused by extensive acceleration in coronal and
interplanetary shock waves (typically associated to fast and energetic
CMEs), which produce harder particle spectra that extend to higher
energies \citep[e.g.][]{Reames2013}. Extreme events can sometimes
produce GeV protons that shower in Earths atmosphere and lead to a
detectable increase in neutron flux at ground level (so called ground
level enhancements (GLEs)).

The spectral and temporal characteristics of solar flares that have
been observed with {\it Fermi} so far can be interpreted in both
scenarios: Some flares show \g-ray emission coincident with the HXR
impulsive emission \citep{Ackermann2012}. The majority of flares,
however, shows longer emission indicative of continuous acceleration
in a stochastic turbulence process in the solar corona and interaction
in the solar upper atmosphere \citep{Ackermann2014}. Alternatively,
continuous acceleration takes place in the shocks of the CME and \g\
rays are produced when particles are travelling along field lines back
to the photosphere and interact there \citep{Ackermann2017}.

One limitation of the LAT and other spacecrafts designed to detect \g\
rays and energetic particles from the sun is their detection area,
which is too small to detect the highest-energy photons, electrons or
ions even in solar flares. Ground-based IACTs, on the other hand, have
sufficient detection area, are operating in the tens of GeV to TeV
energy range, but cannot observe the Sun directly. Extensive air
shower arrays like HAWC or the planned LHAASO detector will be able to
search for solar \g-ray emission in the TeV energy range
\citep{Zhou2017}.

\section{Particle acceleration and \g-ray production in stellar
  flares}\label{sec:stellarflares}

\subsection{Energetics requirements}
\label{sec:energetics}

Next, we will explore if the \g-ray emission from stellar flares can
potentially be detected from nearby active stars with {\it Fermi}-LAT
or IACTs. The fact that no stellar flare from any nearby star has ever
been observed above X-ray energies requires us to combine the
knowledge about solar flares that have been studied in particles and
\g\ rays up to a few GeV, with stellar flares that are typically seen
in radio and optical wavelengths and up to X-ray energies. The
energetics estimates and scaling from the Sun to stellar flares relies
on three main assumptions, which will be discussed in detail:

\begin{enumerate}
\item The underlying physical processes in solar and stellar flares
  are similar. This will allow us to infer the energy in accelerated
  particles in stellar flares in nearby stars.
\item The energy in accelerated electrons is similar to the energy in
  accelerated protons.
\item In stellar superflares, equipartition holds between radiated
  (thermal) X-ray energy and energy in accelerated electrons. This
  allows us to use the measured X-ray flux to infer the energy in
  accelerated protons, independent of the thermal/non-thermal origin
  of X-rays.
\end{enumerate}

\subsubsection{Stellar flares resemble solar flares}

As outlined in \cite{Osten2016}, the interpretation of the \dg\
stellar superflare event assumes that the same physical processes are
at work as in the solar case. In general, this assumption is supported
by many stellar observations, and scaling relations from solar to
stellar flares have been inferred from observations at extreme UV or
soft X-rays (SXR) \citep{Aschwanden2008}. Multi-wavelength
observations show that the inferred properties of the plasma and
non-thermal particles in stellar flares resemble that of solar
flares. For instance, the time evolution of the plasma temperature and
density are similar in solar and stellar flares \citep[see e.g.][and
references therein]{Benz2010}. The most striking correlation in this
regard is probably the linear relation between radio gyro-synchrotron
emission and thermal X-rays, which spans more than 6 orders of
magnitude in X-ray and radio luminosity from solar microflares to the
most extreme stellar flares observed in EQ Peg \citep[c.f. Fig.~11
in][]{Benz2010}.

\subsubsection{Energy in electrons and protons is comparable}

As described in detail in \citet{Benz2010}, ions accelerated in solar
flares impinge on the chromosphere and excite heavy ions. In the
subsequent de-excitation process, nuclear lines are emitted and can be
used to infer the ion spectrum, flux and total energy. Of special
importance is the neutron capture line at 2.223\,MeV. In solar flares,
\citet{Ramaty1995}, for instance, find that the energy in protons with
energies above 1\,MeV is comparable to the energy stored in
non-thermal electrons above 20\,keV. This result is confirmed by
e.g. \citet{Emslie2005}. More recently, also \citet{Emslie2012}
studied the relation between energy released in electrons and ions in
solar flares. The authors found that, for the sample of strong solar
flares that were accompanied by SEPs and CMEs, the energy content in
flare accelerated electrons and ions was comparable. However, the
estimates in these works are afflicted with a considerable uncertainty
and are only valid as order of magnitude estimates mainly due to the
uncertain low-energy cut-off in the electron spectrum and the
extrapolation of the ion spectrum to higher energies. Moreover, not
all strong flares are followed by CMEs and/or SEPs.

\subsubsection{Equipartition holds between thermal and non-thermal
  X-rays}

\citet{Isola2007} find a remarkable correlation between SXR and HXR
radiation from solar flares to the most extreme stellar flares. The
correlation between SXRs (which trace the thermal plasma) and radio
emission (which trace non-thermal electrons) suggests that one can
employ the X-ray emission in stellar flares as a tracer for the energy
content in relativistic electrons and ions, independent of a
thermal/non-thermal origin of the X-rays. Indeed, \citet{Emslie2012}
studied the global energetics of large solar flares and found that the
total SXR radiation from the hot plasma is comparable to the energy
stored in high-energy electrons with energies greater than 20\,keV.

\section{Non-thermal emission model}\label{sec:model}

To produce \g-ray emission at a level detectable by IACTs, two more
conditions have to be fulfilled: Acceleration of particles to VHE
energies with hard \g-ray spectra, and high enough densities for the
production of $\pi^0$-decay \g\ rays. The maximum energy and spectral
index of the relativistic proton population in the \dg\ flare is
probably the most critical and least constrained property in the model
presented below. In analogy to solar flares, we will nevertheless try
to estimate both quantities in two scenarios: Particle acceleration
and \g-ray production in the impulsive phase as a result of magnetic
reconnection, and gradual particle acceleration via diffusive shock
acceleration in the CME that very likely accompanied the flaring
event.

The \dg\ superflare is characterised in optical, radio and X-rays by
two episodes \citep[see][]{Fender2015, Caballero2015, Osten2016}. The
first impulsive episode lasted for approximately 400~seconds and was
denoted ``big first flare'' (BFF). The BFF was followed by a second
flaring episode (called F2) that started around $8\times10^3$ seconds
and peaking at 10$^4$ seconds after the BFF peak. The second flaring
episode lasted for about 6.4\,hours in X-rays \citep{Osten2016} and
$4-5$~days in radio \citep{Fender2015}. The total duration of this
exceptional event was more than two weeks in X-rays.

\subsection{Particle acceleration and \g-ray production in the BFF}

The arguments made in Section~\ref{sec:energetics} suggest that we can
use the measured X-ray emission from stellar flares as a proxy for the
energy content in relativistic protons and nuclei accelerated in the
prompt phase of the \dg\ event. Maximum particle energies associated
with prompt solar flares and as measured by {\it Fermi} reach a few
GeV ($\leq400$\,MeV in \g\ rays) within one minute
\citep{Ackermann2012}. Although particle acceleration in impulsive
solar flares is established, the underlying physical mechanism is not
well understood. Various models exist that link the magnetic
reconnection event more directly or indirectly to the particle
acceleration process \citep[see e.g.][and references
therein]{Knizhnik2011}. For the sake of argument and to guide the
reader, we follow the argumentation of \citet{Takahashi2016}, who
derive a maximum proton energy as $E_{\mathrm{max}} \sim 7$\,GeV
$B_{\mathrm{100}}\,V_{100}\,L_{0.1 \,R_{s}}$. Here, $B_{100}$ is the
active regions (ARs) magnetic field strength in 100\,G, $V_{100}$ is
the ``inflow'' plasma speed in 100\,km\,s$^{-1}$, and $L_{0.1\,R_s}$
is the length scale of the AR in 0.1 solar radii. If equipartition
holds in the \dg\ superflare, the magnetic fields inferred by
\citet{Osten2016} are between a few hundred Gauss and several kG in
the coronal loops, and several kG on the stellar surface. These
magnetic field estimates are consistent with independently inferred
magnetic fields, based on equipartition between gas pressure and
magnetic pressure. Such large magnetic fields would in principle allow
for a fast acceleration of particles similar to solar flares, but to
higher maximum energies than observed for the Sun
\citep{Osten2016}. Assuming their best fit estimate of
$B_{\mathrm{BFF}} = 580$\,G at the site of magnetic reconnection and
an AR length scale of 0.1$R_{s}$ (i.e. 0.25$R_*$), the maximum proton
energy would be around 40\,GeV. Only for somewhat extreme magnetic
fields of several kG and AR sizes comparable to the stellar disk of
\dg\ one would reach maximum energies approaching 1\,TeV. Even if
these extreme conditions could be met, proton spectra in solar flares
have typically indices of $\alpha = 3.0$ or steeper. The best-fit
spectral proton index in a hadronic scenario for the impulsive flare
studied by \citet{Ackermann2012} even has an index of
$\alpha \geq 4.5$.  Nevertheless, we will explore how the \g-ray
spectral energy distribution would look like and how they compare to
the sensitivity of the {\it Fermi}-LAT and the next-generation CTA.

\begin{figure}
  \begin{center}
    \includegraphics[width=0.485\textwidth]{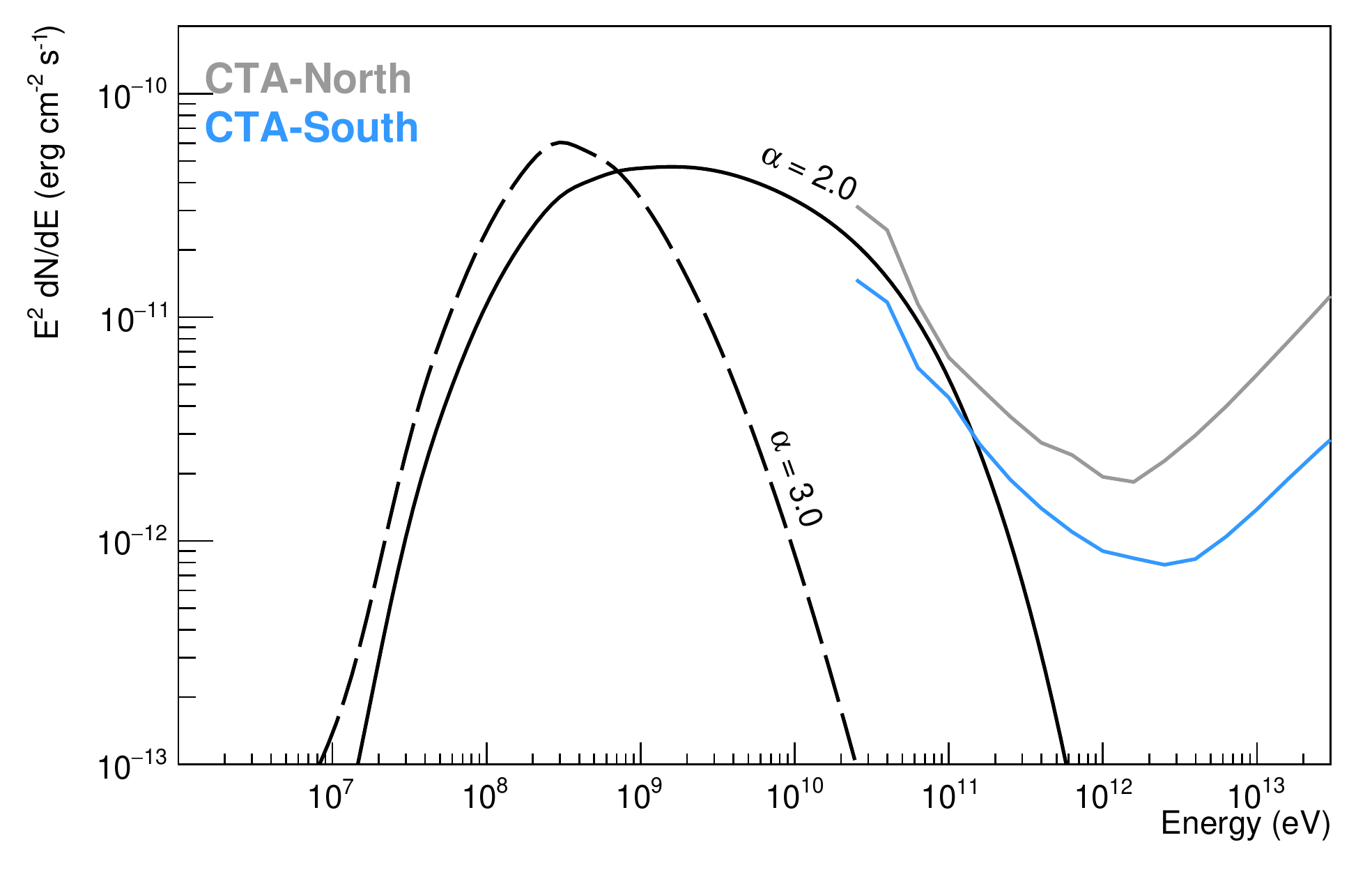}
    \caption{\label{fig1} Expected \g-ray emission from the BFF based
      on the time-dependent injection and interaction model for
      protons as described in the main text. Solid curves depict the
      $\pi^0$-decay \g-ray emission from protons with cut-off energy
      $E_c = 1$\,GeV and index $\alpha=2.0$. Dashed curves show the
      emission for protons with $E_c = 100$\,GeV and
      $\alpha=3.0$. Also shown are the CTA-South and CTA-North
      integral sensitivities for the duration of the flare of 2-hours
      for the BFF. The 1-day {\it Fermi}-LAT upper limit from
      \citet{Loh2017} would be around
      $2.7\times 10^{-9}$\,erg\,cm$^{-2}$\,s$^{-1}$, assuming a mean
      photon energy of 3\,GeV.}
  \end{center}
\end{figure}

As discussed in Section~\ref{sec:solarflares}, observations of solar
flares with {\it Fermi} show that the high-energy \g-ray emission can
last for hours, although the short impulsive phase typically seen at
X-ray energies only lasts for seconds to minutes \citep[see e.g.][and
references therein]{Ackermann2014}. Given the characteristics of
impulsive solar flares, we explore two cases here that could be seen
as realistic and optimistic. Injection of protons according to a
power-law in energy with exponential cut-off of the form:
$$dN(E)/dE = N_0E^{-\alpha}\exp(-E/E_c)~,$$ within 2 hours in
the BFF with a minimum energy of $E_{p,\mathrm{min}} = 1$\,GeV and 1)
with a cut-off energy of $E_c = 50$\,GeV and $\alpha=3.0$; and 2) with
a cut-off energy of $E_c = 500$\,GeV and $\alpha=2.0$. In both cases,
the energy in injected protons corresponds to the total energy in
electrons as inferred by \citet{Osten2016} based on the {\it Swift}
data between $0.3-10$\,keV. As described above, we assume energy
partition between electrons and protons and hence adopt the electron
number density in the coronal loop as inferred by \citet{Osten2016} of
$N_e = 3\times 10^{11}$\,cm$^{-3}$ as target density for $pp$
interactions. In all considerations we adopt a distance to \dg\ of
18\,pc. Figure~\ref{fig1} shows the resulting $\pi^0$-decay \g-ray
spectra of the two BFF scenarios in a time-dependent model for the
injection and interaction of relativistic protons based on the
radiative code by \citet{Hinton2007}. Also shown is the expected CTA
sensitivity\footnote{Integral energy flux sensitivities have been
  obtained using the public CTA performance files
  (\url{https://www.cta-observatory.org/science/cta-performance/}),
  ignoring systematic errors and assuming a background normalisation
  value of $\alpha=0.05$.}.  Interestingly, in a high-density
environment such as the one considered here, the cooling time of
protons is comparable to their injection time, i.e.
$\tau_{pp} \simeq 1.6$\,hrs. This implies that, independent of maximum
energy and energy content in accelerated particles, a significant
fraction of the energy injected in protons is radiated at \g-ray
energies and not released into interplanetary space as stellar
energetic particles.

It is clear that only in the optimistic scenario \g-ray emission at a
level detectable by CTA is produced. Reducing the minimum energy for
the injected protons from 1\,GeV to 100\,MeV would reduce the level of
\g-ray emission in the extreme scenario by $\sim 25\%$. Note that the
1-day LAT upper limit by \citet{Loh2017} is about an order of
magnitude higher than the predicted emission level at any energy.

\subsection{Particle acceleration and \g-ray production in the F2}

Energetic particles released in solar flares are often accelerated to
hundreds of MeV and GeV energies. IACTs, however, have an energy
threshold in the tens of GeV energy range, requiring particle
acceleration to hundreds of GeV up to TeV energies in stellar
flares. The highest-energy particles and GLE events, are often
associated with massive and fast CMEs
\citep[e.g.][]{Gopalswamy2012}. Although faster CMEs tend to produce
higher-energy particles more efficiently, there is a large scatter of
SEP intensities released in these gradual events, independent of CME
properties. There is increasing evidence that the most energetic SEP
and GLE events are associated with a population of supra-thermal
particles pre-accelerated in preceding large flares
\citep[e.g.][]{Reames2000, Kahler2001}, often from the same AR
\citep{Gopalswamy2004}. Generally, quasi-parallel shocks are able to
accelerate more protons and nuclei out of the thermal pool of
solar/stellar wind particles, but not to sufficiently high energies
and typically with steep spectra. Quasi-perpendicular shocks, on the
other hand, are able to accelerate particles to higher energies faster
and produce harder particle spectra. However, injection into the shock
region is hampered as it is harder for particles to catch up with the
shock and enter the acceleration process \citep{Kozarev2016}. A
supra-thermal seed particle population allows for both, an efficient
injection into quasi-perpendicular shocks and fast acceleration to
high energies \citep[e.g.][]{Ng2008, Laming2013}. In fact, the extreme
solar flare from January 20$^{\mathrm{th}}$, 2005 is a good example to
support this hypothesis: This event was preceded by several X-class
solar flares from the same AR and the resulting final SEP spectrum had
a hard particle index of $\alpha=2.14$ \citep{Mewaldt2012}.

\subsubsection{Energy in relativistic protons in CME shock}
\label{sec:F2energy}
The strongest solar flares are associated almost 1:1 with fast and
equally energetic CMEs \citep{Emslie2012}. The direct imaging of
stellar CMEs is implausible with current instruments, why indirect
measures are needed to detect stellar CMEs and to infer their
properties, such as speed or mass loss \citep[e.g.][and references
therein]{Leitzinger2011, Kay2016}. In the following we will assume
that the F2 event in \dg\ is accompanied by a CME. Using the findings
of \citet{Emslie2012} again, the kinetic energy in a CME is
approximately an order of magnitude larger than the energy in
non-thermal electrons and ions. The energy in SEPs on the other hand,
is about 5\% of the CME kinetic energy. For the sake of argument and
given the large uncertainties, we will in the following assume that
the energy in SEP is the same as the energy in flare-accelerated
electrons, i.e. $9\times10^{35}$\,erg for F2. We base our estimates of
the dynamics of the CME in F2, i.e. the mass lost during the event and
the speed of the ejecta, on the empirical relation between flare
energy and mass-loss in T Tauri star superflares and solar flares as
discussed by \citet{Aarnio2012} (cf. their Fig.~1). For a kinetic
energy of $E_{\rm CME, kin} = 9\times10^{36}$\,erg, eqn. 2 of
\citet{Aarnio2012} gives an ejected mass in the range of
$m_{\rm CME} = (4.9\times10^{19} - 3.8\times10^{21})$ grams. Guided by
the case for solar flares \citep[e.g.][]{Mewaldt2012}, we will adopt
here a CME velocity of $v_{\rm CME} = 3000$\,km\,s$^{-1}$, which gives
an ejected mass of $m_{\rm CME} = 1.8\times10^{20}$ grams -- in
agreement with \citet{Aarnio2012}.

\subsubsection{Maximum energy of accelerated particles}
Estimating the maximum particle energy that can be achieved in the CME
shock acceleration process is challenging, but can again be based on
solar flares and associated CME acceleration. Simulations of particle
acceleration in parallel shocks show that particle energies of
hundreds of MeV energies can be achieved within 10 minutes in a CME of
2500\,km\,s$^{-1}$ \citep{Ng2008}. \citet{Kozarev2016}, on the other
hand, developed a data-driven analytical model, and find that with
slower shock speeds of 800\,km\,s$^{-1}$ and acceleration in
quasi-perpendicular shocks, maximum energies of several GeV can be
reached in the same time. Scaling these results to the duration of the
\dg\ flare of 10\,hrs, and ignoring non-linear effects such as
proton-amplified Alfv\'en wave growth, suggests that maximum particle
energies of 20 GeV in the case of a fast and quasi-parallel shock, and
100\,GeV in case of a quasi-perpendicular shock can be reached. In the
\citet{Kozarev2016} model, a constant magnetic field of 5\,G was
assumed. In reality, acceleration in the CME shock will proceed under
changing conditions and the magnetic field will drop as the CME
expands as $R^{-2}$ \citep{Kay2016}. An assumed 20\,kG field --
typical for an M-dwarf AR magnetic field -- will have dropped to
$\sim20$\,G after 2 hours, and to about $\sim1$\,G after 10 hours for
the assumed shock speed of $v_{\rm CME} = 3000$\,km\,s$^{-1}$. Since
the detailed modelling of particle acceleration under changing
magnetic field conditions is beyond the scope of this paper, we will
assume that acceleration proceeds in a constant magnetic field of
10\,G. If indeed strong shocks are present in the CME, maximum
particle energies of 200\,GeV for fast parallel shocks and 1\,TeV for
somewhat slower quasi-perpendicular shocks are possible. The real
conditions under which particle acceleration in the CME proceeds is
much more complex, as e.g. the shock morphology will change from
quasi-perpendicular to quasi-parallel as the CME expands. As discussed
above, if the origin of the two \dg\ flares was indeed the same AR,
the coronal and ejected particles from the BFF, will have very likely
been pre-accelerated. This would result in a more efficient injection
into the F2 CME shock and acceleration to higher particle energies and
with hard spectra. To summarise, particle acceleration in the fast and
dense CME that was likely associated to the \dg\ F2 event to energies
beyond 100\,GeV and possibly up to TeV energies seems plausible.

The maximum energy particles can be accelerated to is limited either
by the acceleration time (i.e. the duration of the flare), or by the
$pp$ loss time. The acceleration time $\tau_{\rm acc}$ in a strong
shock and in the Bohm limit (i.e. the theoretically fastest possible
acceleration) is:
$$\tau_{\rm acc} \approx 1.4\,{\rm hrs}\,v^{-2}_{\rm CME,
  10^3\,{km\,s^{-1}}}\,E_{\rm 100\,GeV}\,B^{-1}_{\rm G}.$$
The $pp$ loss time is \citep[e.g.][]{Hinton2009}:
$$\tau_{pp} \approx 2.6\,{\rm hrs}\,n_{11}^{-1},$$
with $n$ being the target density in $10^{11}$\,cm$^{-3}$.
Especially at early times, the $pp$ cooling time is short, as the
density in the CME is very high. As the density drops as
$\rho_{\rm CME} \propto t^{-3}$ and the magnetic field only falls like
$B\propto t^{-2}$, the maximum energy of particles will increase over
the duration of the flare. Once the $pp$ cooling time becomes shorter
than the duration of the flare, the maximum particle energy will drop
linearly over time again. For an initial magnetic field in the AR of
$B_0 = 5\,kG$, and a CME shock speed of
$v_{\rm CME} = 3000\,$km\,s$^{-1}$, maximum particle energies will
increase from 100\,GeV after 15 minutes to about 6\,TeV after 5 hours,
and drop to 1.5\,TeV after 10 hours. These estimates should not be
taken at face value: they rely on the assumption of a strong shock and
$pp$ interaction in a homogeneous medium. If acceleration operates in
weaker shocks, lower maximum energies are to be expected. On the other
hand, if only a fraction of protons interact downstream of the shock,
the bulk can be accelerated to higher energies earlier on.

\subsubsection{Expected \g-ray emission from the F2
  event} \label{sec:F2emission}
In the following we will explore different scenarios to give a rough
overview over the expected \g-ray emission under varying (but
constant) conditions. We will close this chapter by exploring a
time-dependent scenario that could be seen as representative for the
conditions in the F2 flare in this model.

\begin{figure*}
  \begin{center}
    \includegraphics[width=0.485\textwidth]{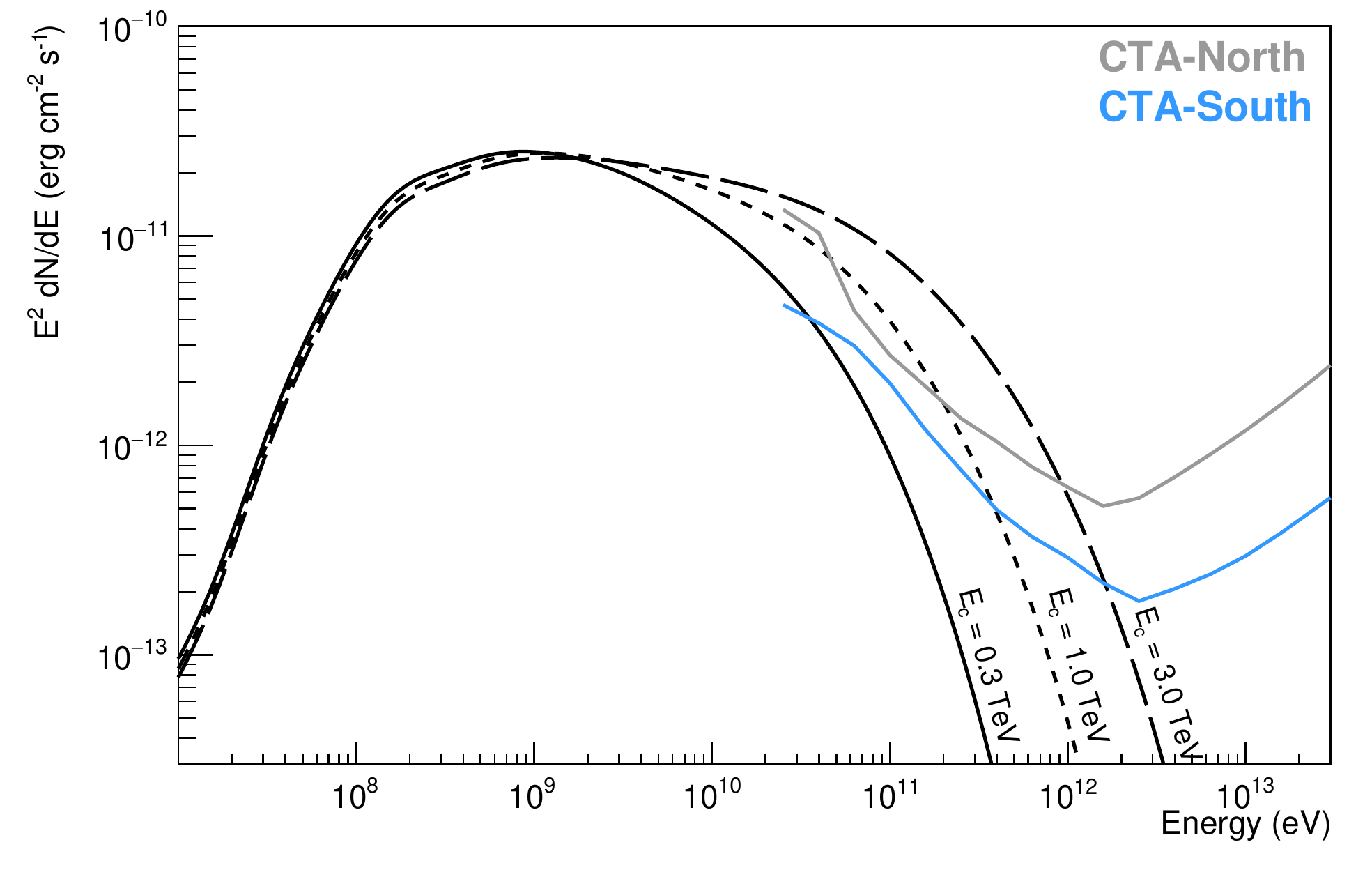}
    \includegraphics[width=0.485\textwidth]{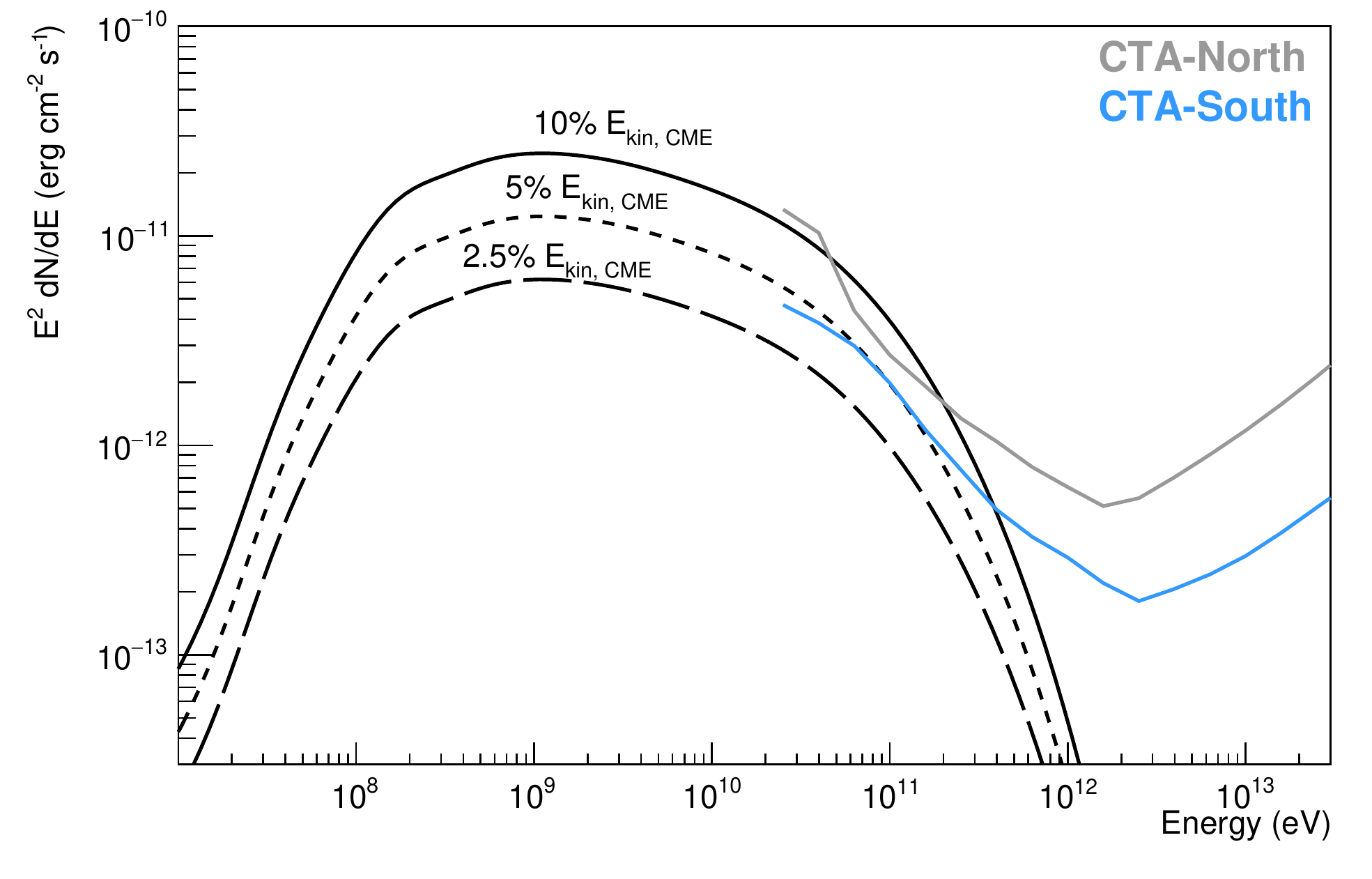}
    \caption{\label{fig2} Expected \g-ray emission of the F2 in the
      non-thermal emission model as described in the main text and in
      Fig.~\ref{fig1}. Shown are models with varying proton cutoff
      energy (left, $E_{\mathrm{inj}} = 9\times10^{35}$\,erg) and
      fractions of CME kinetic energy that are transferred to
      high-energy protons (right, $E_c = 1$\,TeV). In all models an
      average density of $n_{\rm CME} = 1.6\times 10^{11}$, and
      injection over a time of 10 hours is assumed. Also shown are the
      integral energy flux sensitivities for the two proposed CTA
      sites for an exposure of 10 hours.}
  \end{center}
\end{figure*}

As discussed in Section~\ref{sec:F2energy}, it is assumed that
material with a mass of $m_{\rm CME} = 1.8\times10^{20}$\,g is ejected
in the CME and moving with a speed of
$v_{\rm CME} = 3000$\,km\,s$^{-1}$. With this assumed CME velocity and
mass, the density would drop after 5 hours to
$n_{\rm CME} = 1.6\times 10^{11}$, and particles would start to leave
the system, as the $pp$ cooling time would be comparable to the flare
duration. Protons are injected constantly over the duration of the F2
event and according to a power-law in energy with exponential cutoff
of the form:
$$dN(E)/dE = N_0E^{-\Gamma}\exp(-E/E_c)~,$$ with proton spectral index
$\Gamma=2.0$ and cutoff energy $E_c=1$\,TeV. In a first attempt to
study the expected emission from F2, all parameters are fixed and only
the total energy in protons is varied. Figure~\ref{fig2} shows the
expected \g-ray emission for three proton cut-off energies, and for
three different injection energies, while leaving the other quantity
at a fixed value.

In this simplified model, \g-ray emission from an event similar to the
\dg\ flare would be visible for the F2 episode with both CTA-North as
well as CTA-South for events where 10\% of the potential CME kinetic
energy can be transferred to accelerated protons, and if protons are
accelerated to energies higher than 1\,TeV. If the maximum proton
energy reaches 300\,GeV, this event would still be visible with
CTA-South. For a fixed maximum proton energy of 1\,TeV, on the other
hand, the F2 event would be visible from both CTA sites if
$\gtrsim$7\% of the CME kinetic energy are transferred to accelerated
protons. As discussed in \citet{Emslie2012}, about 5\% of the CME
kinetic energy is transferred to SEPs in solar flares. In this case,
the F2 event would be visible with CTA-South, but not CTA-North. In
all scenarios, a low energy threshold is key to study the potential
\g-ray emission in the high-energy cut-off region.

\begin{table}
  \centering
  \begin{tabular}{ccccc}
    \hline
    \hline
    Phase & Duration & $E_{\rm pp}$ & $E_{\rm max}$ & $\rho$ \\
          & hrs & $10^{35}$\,erg & TeV & $10^{11}$\,cm$^{-3}$ \\ \hline
    Early & 3.0 & 2.7 & 0.3 & 30 \\
    Mid   & 4.0 & 3.6 & 1.0 & 1.5 \\
    Late  & 3.0 & 2.7 & 0.8 & 0.5 \\
    \hline
    \hline
  \end{tabular}
  \caption{Summary of input parameters for the time-dependent
    injection model for the three representative phases of F2
    as discussed in the main text and shown in Fig.~\ref{fig3}.}
  \label{tab1}
\end{table}

\begin{figure}
    \includegraphics[width=0.485\textwidth]{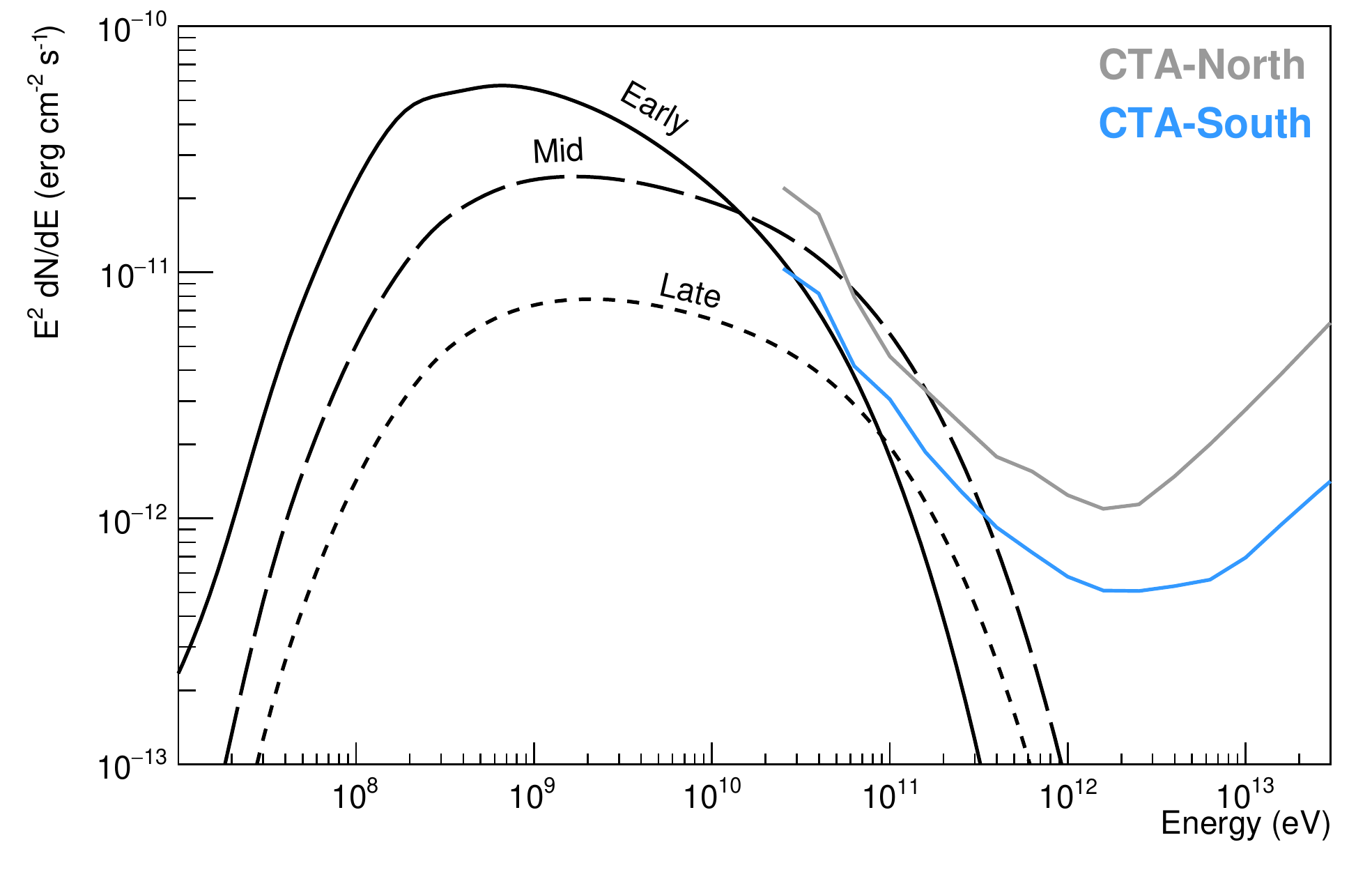}
    \caption{\label{fig3} Expected \g-ray emission of the three
      representative F2 in the non-thermal emission model as described
      in the main text and in Table~\ref{tab1}. Also shown are the
      integral energy flux sensitivities for the two proposed CTA
      sites for an assumed observation time of 4 hours.}
\end{figure}

As discussed in Section \ref{sec:F2energy} and the beginning of this
section, conditions such as the magnetic field, target density and
maximum particle energy change as the CME moves outwards. Next we will
try to address this by exploring different environmental conditions,
that could be seen as representative for the early, mid and late
phases of the F2 event. Table~\ref{tab1} summarises the assumed input
parameters and Figure~\ref{fig3} shows the expected \g-ray emission
for such a model. The early phase can be characterised by acceleration
in a very dense medium with somewhat lower maximum energies. It is
followed by a phase, where the $pp$ cooling time is comparable to the
flare duration and particles start to leave the system. Here, the
highest maximum energies are expected to be achieved in the presented
model. At later times, the maximum particle energy will drop, as the
magnetic field decays. As the CME material is thinning out over time,
densities are falling below a critical value, where no more detectable
\g-ray emission is expected and the bulk of accelerated particles is
assumed to leave the system as stellar energetic
particles. Interestingly, planets situated in the habitable zone
around M-dwarf stars are very close to the central star
\citep[$\sim$0.1\,AU,][]{Kay2016}. This implies that for such extreme
events as the one seen from \dg\ and the shock speeds assumed above,
the CME will impact planets situated in the habitable zone while the
system is still calorimetric (i.e. all accelerated protons will
interact). For the extreme (and more frequent) events, densities
should drop below the calorimetric limit before reaching the habitable
zone. In reality, the situation is much more complex and a fraction of
the accelerated particles will escape ahead of the shock, while others
will stay in the acceleration process or interact downstream with the
released CME material producing $\pi^0$-decay \g\ rays.

\section{Expectation for the population of nearby young stars}
\label{sec:cta}

\begin{table*}
  \centering
  \begin{tabular}{ccccccc}
    \hline
    \hline
    Object & Distance & Flare Energy & Fluence & Energy range & Duration & Reference\\
           & pc & 10$^{34}$\,erg & 10$^{-5}$\,erg\,cm$^{-2}$ & & hours & \\ \hline
    AD Leo & 4.9 & 1.0 & 0.35 & $(1800 - 8000)$\,\AA\ & 4.0 & \citet{Hawley1991}\\
    II Peg & 42.0 & 600 & 2.9 & $(0.01 - 200)$\,keV & $>$3.3 & \citet{Osten2007}\\
    EV Lac & 5.0 & 5.8 & 1.9 & $(0.3 - 10)$\,keV & 1.7 & \citet{Osten2010,Osten2016}\\
    \dg\ & 18.0 & (40 + 90)$^\dagger$ & 3.4 & $(0.3 - 10)$\,keV & $>$8 & \citet{Osten2016}\\
    EQ Peg & 6.5 & 9.0 & 1.8 & $(2.0 - 20)$\,keV & 1.1 & \citet{Tsuboi2016}\\
    AT Mic & 10.2 & 30 & 2.4 & $(2.0 - 20)$\,keV & 1.7 & \citet{Tsuboi2016}\\
    YZ CMi & 5.9 & $<$20 & $<$4.8 & $(2.0 - 20)$\,keV & $<$1.5 & \citet{Tsuboi2016}\\
    \hline
    \hline
  \end{tabular}
  \newline $^\dagger$ Energy released in BFF and F2.
  \caption{List of nearby active stars and recorded superflares, their
    energetics, duration and fluence.}
  \label{tab:flares}
\end{table*}

Stellar superflares are exceptional events and only a handful have
been studied in detail so far. Table~\ref{tab:flares} summarises the
studied superflares and their properties. In the model described
above, and based on the properties such as fluence and duration, all
superflares would have been within reach of CTA. In a systematic
search for flares with the MAXI/GSC instrument aboard the ISS,
\citet{Tsuboi2016} found only 13 stars to be flaring out of a
population of 256 active binary stars \citep{Eker2008}. Some of these
stars flared multiple times and experienced radiative energy releases
of up to 10$^{39}$ ergs. Given the infrequency of these large and
energetic flares that would potentially be visible for Cherenkov
telescopes, the occurence rate is rather hard to
predict. \citet{Osten2016} estimated flares with U-band energy
releases similar to the \dg\ event to happen about once every 5 months
for YZ CMi and EV Lac, and once every 3 years to happen on EQ Peg,
respectively. $(10 - 15)$\%, which reduces the number of flares that
are possible to follow-up on promptly. Although the frequency of these
superflares is rather uncertain, about one event per year should be
energetic enough such that observations with MAGIC, H.E.S.S. or
VERITAS\footnote{Assuming that the sensitivity of current generation
  instruments is a factor 5-10 worse than CTA-South.} should be able
to constrain the here proposed model, the energy in protons and the
maximum proton energy.

\section{Conclusions and Outlook}

We have estimated the \g-ray emission from flaring, nearby, young
M-dwarf stars using the X-ray properties of the \dg\ superflare and
extrapolations from solar flares. In the considered scenario, \g-ray
emission from the \dg\ event in the BFF at a level detectable by the
future CTA is possible if equipartition holds between non-thermal
electrons and ions, and if particles are accelerated to TeV energies
with hard particle spectra. Hard particle spectra and acceleration to
TeV energies seems to be challenging at least in this impulsive phase
of the flare. \g-ray emission in gradual solar flares and associated
SEPs have been detected up to several GeV, implying acceleration of
particles to tens of GeV energies in the CMEs. We argued that particle
acceleration in the F2 event of the \dg\ flare could be realised and
that acceleration to high energies and with hard particle indices are
plausible. Considering the extreme conditions in young M-dwarf stars
such as their fast rotation periods, strong magnetic fields and
energetic eruptions compared to the Sun, it is not unrealistic to
assume that particles are accelerated to hundreds of GeV energies in
M-dwarf flares. If detected, the discovery of \g-ray emission from a
single star would represent a major breakthrough in stellar physics
and mark the detection of GeV \g\ rays from the first stellar object
other than the Sun. It would also give insights into particle
acceleration and interaction in stellar atmospheres. Furthermore, it
would give a handle on the cosmic-ray flux near young stars that are
often considered to host exoplanets but experience much stronger
flares than observed in the Sun. Studies suggest that cosmic-rays
accelerated in stellar flares can have a significant astrobiological
effect on exoplanetary atmospheres. A flare from the star system AD
Leo for instance (which was 20 times fainter than the \dg\ flare), is
estimated to have removed the ozone layer of a potential exoplanet
over the course of two years, with a recovery time of several decades
\citep{Segura2010}. Given the occurrence rate of these strong flares,
there is a possibility of a permanent erosion of the ozone layer in
exoplanets \citep{Osten2016}.

Large field-of-view optical survey instruments such as the Zwicky
Transient Factory (ZTF) or the Large Synoptic Survey Telescope (LSST)
will detect an increasing number of strong stellar flares and add to
our understanding of the physics of stellar flares and their impact on
the habitability of planets. At very high energies, CTA could
potentially open a new window for the study of non-thermal particles
in stellar flares.

\section*{Acknowledgements}
We thank the anonymous referee for her/his comments, which
considerably improved the quality of the paper. We'd also like to
thank M. Ackermann, R. B\"uhler, J.A. Hinton, G. Maier, \L. Stawarz
and A. Taylor for fruitful discussions and careful reading of the
manuscript, as well as K. Strotjohann for contributions to an early
version of the draft. This paper has gone through internal review by
the CTA Consortium.

\bibliographystyle{mn2e_williams}
\bibliography{FlaringStars_paper}

\begin{thebibliography}{42}
\expandafter\ifx\csname natexlab\endcsname\relax\def\natexlab#1{#1}\fi

\bibitem[{{Aarnio} {et~al}\mbox{.}(2012){Aarnio}, {Matt}, \&
  {Stassun}}]{Aarnio2012}
{Aarnio} A.~N., {Matt} S.~P., {Stassun} K.~G., 2012, \apj, 760, 9

\bibitem[{{Abdo} {et~al}\mbox{.}(2011){Abdo}, {Ackermann}, {Ajello}, {Baldini},
  {Ballet}, {Barbiellini}, {Bastieri}, {Bechtol}, {Bellazzini}, {Berenji},
  {Bonamente}, {Borgland}, {Bouvier}, {Bregeon}, {Brez}, {Brigida}, {Bruel},
  {Buehler}, {Buson}, {Caliandro}, {Cameron}, {Caraveo}, {Casandjian},
  {Cecchi}, {Charles}, {Chekhtman}, {Chiang}, {Ciprini}, {Claus},
  {Cohen-Tanugi}, {Conrad}, {Cutini}, {de Angelis}, {de Palma}, {Dermer},
  {Digel}, {Silva}, {Drell}, {Dubois}, {Favuzzi}, {Fegan}, {Focke}, {Fortin},
  {Frailis}, {Funk}, {Fusco}, {Gargano}, {Gasparrini}, {Gehrels}, {Germani},
  {Giglietto}, {Giordano}, {Giroletti}, {Glanzman}, {Godfrey}, {Grenier},
  {Grillo}, {Guiriec}, {Hadasch}, {Hays}, {Hughes}, {Iafrate},
  {J{\'o}hannesson}, {Johnson}, {Johnson}, {Kamae}, {Katagiri}, {Kataoka},
  {Kn{\"o}dlseder}, {Kuss}, {Lande}, {Latronico}, {Lee}, {Lionetto}, {Longo},
  {Loparco}, {Lott}, {Lovellette}, {Lubrano}, {Makeev}, {Mazziotta}, {McEnery},
  {Mehault}, {Michelson}, {Mitthumsiri}, {Mizuno}, {Moiseev}, {Monte},
  {Monzani}, {Morselli}, {Moskalenko}, {Murgia}, {Nakamori}, {Naumann-Godo},
  {Nolan}, {Norris}, {Nuss}, {Ohsugi}, {Okumura}, {Omodei}, {Orlando}, {Ormes},
  {Ozaki}, {Paneque}, {Pelassa}, {Pesce-Rollins}, {Pierbattista}, {Piron},
  {Porter}, {Rain{\`o}}, {Rando}, {Razzano}, {Reimer}, {Reimer}, {Reposeur},
  {Ritz}, {Sadrozinski}, {Schalk}, {Sgr{\`o}}, {Share}, {Siskind}, {Smith},
  {Spandre}, {Spinelli}, {Strickman}, {Strong}, {Takahashi}, {Tanaka},
  {Thayer}, {Thayer}, {Thompson}, {Tibaldo}, {Torres}, {Tosti}, {Tramacere},
  {Troja}, {Uchiyama}, {Usher}, {Vandenbroucke}, {Vasileiou}, {Vianello},
  {Vilchez}, {Vitale}, {Vladimirov}, {Waite}, {Wang}, {Winer}, {Wood}, {Yang},
  \& {Ziegler}}]{Abdo2011}
{Abdo} A.~A. {et~al.}, 2011, \apj, 734, 116

\bibitem[{{Ackermann} {et~al}\mbox{.}(2014){Ackermann}, {Ajello}, {Albert},
  {Allafort}, {Baldini}, {Barbiellini}, {Bastieri}, {Bechtol}, {Bellazzini},
  {Bissaldi}, {Bonamente}, {Bottacini}, {Bouvier}, {Brandt}, {Bregeon},
  {Brigida}, {Bruel}, {Buehler}, {Buson}, {Caliandro}, {Cameron}, {Caraveo},
  {Cecchi}, {Charles}, {Chekhtman}, {Chen}, {Chiang}, {Chiaro}, {Ciprini},
  {Claus}, {Cohen-Tanugi}, {Conrad}, {Cutini}, {D'Ammando}, {de Angelis}, {de
  Palma}, {Dermer}, {Desiante}, {Digel}, {Di Venere}, {Silva}, {Drell},
  {Drlica-Wagner}, {Favuzzi}, {Fegan}, {Focke}, {Franckowiak}, {Fukazawa},
  {Funk}, {Fusco}, {Gargano}, {Gasparrini}, {Germani}, {Giglietto}, {Giordano},
  {Giroletti}, {Glanzman}, {Godfrey}, {Grenier}, {Grove}, {Guiriec}, {Hadasch},
  {Hayashida}, {Hays}, {Horan}, {Hughes}, {Inoue}, {Jackson}, {Jogler},
  {J{\'o}hannesson}, {Johnson}, {Kamae}, {Kawano}, {Kn{\"o}dlseder}, {Kuss},
  {Lande}, {Larsson}, {Latronico}, {Lemoine-Goumard}, {Longo}, {Loparco},
  {Lott}, {Lovellette}, {Lubrano}, {Mayer}, {Mazziotta}, {McEnery},
  {Michelson}, {Mizuno}, {Moiseev}, {Monte}, {Monzani}, {Moretti}, {Morselli},
  {Moskalenko}, {Murgia}, {Murphy}, {Nemmen}, {Nuss}, {Ohno}, {Ohsugi},
  {Okumura}, {Omodei}, {Orienti}, {Orlando}, {Ormes}, {Paneque}, {Panetta},
  {Perkins}, {Pesce-Rollins}, {Petrosian}, {Piron}, {Pivato}, {Porter},
  {Rain{\`o}}, {Rando}, {Razzano}, {Reimer}, {Reimer}, {Ritz}, {Schulz},
  {Sgr{\`o}}, {Siskind}, {Spandre}, {Spinelli}, {Takahashi}, {Takeuchi},
  {Tanaka}, {Thayer}, {Thayer}, {Thompson}, {Tibaldo}, {Tinivella}, {Tosti},
  {Troja}, {Tronconi}, {Usher}, {Vandenbroucke}, {Vasileiou}, {Vianello},
  {Vitale}, {Werner}, {Winer}, {Wood}, {Wood}, {Wood}, {Yang}, \& {Fermi LAT
  Collaboration}}]{Ackermann2014}
{Ackermann} M. {et~al.}, 2014, \apj, 787, 15

\bibitem[{{Ackermann} {et~al}\mbox{.}(2012){Ackermann}, {Ajello}, {Allafort},
  {Atwood}, {Baldini}, {Barbiellini}, {Bastieri}, {Bechtol}, {Bellazzini},
  {Bhat}, {Blandford}, {Bonamente}, {Borgland}, {Bregeon}, {Briggs}, {Brigida},
  {Bruel}, {Buehler}, {Burgess}, {Buson}, {Caliandro}, {Cameron}, {Casandjian},
  {Cecchi}, {Charles}, {Chekhtman}, {Chiang}, {Ciprini}, {Claus},
  {Cohen-Tanugi}, {Connaughton}, {Conrad}, {Cutini}, {Dennis}, {de Palma},
  {Dermer}, {Digel}, {Silva}, {Drell}, {Drlica-Wagner}, {Dubois}, {Favuzzi},
  {Fegan}, {Ferrara}, {Fortin}, {Fukazawa}, {Fusco}, {Gargano}, {Germani},
  {Giglietto}, {Giordano}, {Giroletti}, {Glanzman}, {Godfrey}, {Grillo},
  {Grove}, {Gruber}, {Guiriec}, {Hadasch}, {Hayashida}, {Hays}, {Horan},
  {Iafrate}, {J{\'o}hannesson}, {Johnson}, {Johnson}, {Kamae}, {Kippen},
  {Kn{\"o}dlseder}, {Kuss}, {Lande}, {Latronico}, {Longo}, {Loparco}, {Lott},
  {Lovellette}, {Lubrano}, {Mazziotta}, {McEnery}, {Meegan}, {Mehault},
  {Michelson}, {Mitthumsiri}, {Monte}, {Monzani}, {Morselli}, {Moskalenko},
  {Murgia}, {Murphy}, {Naumann-Godo}, {Nuss}, {Nymark}, {Ohno}, {Ohsugi},
  {Okumura}, {Omodei}, {Orlando}, {Paciesas}, {Panetta}, {Parent},
  {Pesce-Rollins}, {Petrosian}, {Pierbattista}, {Piron}, {Pivato}, {Poon},
  {Porter}, {Preece}, {Rain{\`o}}, {Rando}, {Razzano}, {Razzaque}, {Reimer},
  {Reimer}, {Ritz}, {Sbarra}, {Schwartz}, {Sgr{\`o}}, {Share}, {Siskind},
  {Spinelli}, {Takahashi}, {Tanaka}, {Tanaka}, {Thayer}, {Tibaldo},
  {Tinivella}, {Tolbert}, {Tosti}, {Troja}, {Uchiyama}, {Usher},
  {Vandenbroucke}, {Vasileiou}, {Vianello}, {Vitale}, {von Kienlin}, {Waite},
  {Wilson-Hodge}, {Wood}, {Wood}, {Yang}, \& {Fermi LAT
  Collaboration}}]{Ackermann2012}
{Ackermann} M. {et~al.}, 2012, \apj, 745, 144

\bibitem[{{Ackermann} {et~al}\mbox{.}(2017){Ackermann}, {Allafort}, {Baldini},
  {Barbiellini}, {Bastieri}, {Bellazzini}, {Bissaldi}, {Bonino}, {Bottacini},
  {Bregeon}, {Bruel}, {Buehler}, {Cameron}, {Caragiulo}, \&
  {Caraveo}}]{Ackermann2017}
{Ackermann} M. {et~al.}, 2017, \apj, 835, 219

\bibitem[{{Ajello} {et~al}\mbox{.}(2014){Ajello}, {Albert}, {Allafort},
  {Baldini}, {Barbiellini}, {Bastieri}, {Bellazzini}, {Bissaldi}, {Bonamente},
  {Brandt}, {Bregeon}, {Brigida}, {Bruel}, {Buehler}, {Buson}, {Caliandro},
  {Cameron}, {Caraveo}, {Cecchi}, {Charles}, {Chekhtman}, {Chiang}, {Chiaro},
  {Ciprini}, {Claus}, {Cohen-Tanugi}, {Cominsky}, {Conrad}, {Cutini},
  {D'Ammando}, {de Palma}, {Dermer}, {Desiante}, {Digel}, {Silva}, {Drell},
  {Drlica-Wagner}, {Favuzzi}, {Focke}, {Franckowiak}, {Fukazawa}, {Fusco},
  {Gargano}, {Gasparrini}, {Germani}, {Giglietto}, {Giommi}, {Giordano},
  {Giroletti}, {Glanzman}, {Godfrey}, {Grenier}, {Grove}, {Guiriec}, {Hadasch},
  {Hayashida}, {Hays}, {Horan}, {Hou}, {Hughes}, {Inoue}, {Jackson}, {Jogler},
  {J{\'o}hannesson}, {Johnson}, {Johnson}, {Kamae}, {Kn{\"o}dlseder},
  {Kocevski}, {Kuss}, {Lande}, {Larsson}, {Latronico}, {Longo}, {Loparco},
  {Lott}, {Lovellette}, {Lubrano}, {Mayer}, {Mazziotta}, {McEnery},
  {Michelson}, {Mizuno}, {Moiseev}, {Monte}, {Monzani}, {Morselli},
  {Moskalenko}, {Murgia}, {Murphy}, {Nakamori}, {Nemmen}, {Nuss}, {Ohno},
  {Ohsugi}, {Omodei}, {Orienti}, {Orlando}, {Ormes}, {Paneque}, {Panetta},
  {Perkins}, {Pesce-Rollins}, {Petrosian}, {Piron}, {Pivato}, {Porter},
  {Rain{\`o}}, {Rando}, {Razzano}, {Reimer}, {Reimer}, {Roth}, {Schulz},
  {Sgr{\`o}}, {Siskind}, {Spandre}, {Spinelli}, {Takahashi}, {Thayer},
  {Thayer}, {Thompson}, {Tibaldo}, {Tinivella}, {Tosti}, {Troja}, {Usher},
  {Vandenbroucke}, {Vasileiou}, {Vianello}, {Vitale}, {Werner}, {Winer},
  {Wood}, {Wood}, \& {Yang}}]{Ajello2014}
{Ajello} M. {et~al.}, 2014, \apj, 789, 20

\bibitem[{{Aschwanden} {et~al}\mbox{.}(2008){Aschwanden}, {Stern}, \&
  {G{\"u}del}}]{Aschwanden2008}
{Aschwanden} M.~J., {Stern} R.~A., {G{\"u}del} M., 2008, \apj, 672, 659

\bibitem[{{Benz} \& {G{\"u}del}(2010)}]{Benz2010}
{Benz} A.~O., {G{\"u}del} M., 2010, \araa, 48, 241

\bibitem[{{Caballero-Garc{\'{\i}}a}
  {et~al}\mbox{.}(2015){Caballero-Garc{\'{\i}}a}, {{\v S}imon},
  {Jel{\'{\i}}nek}, {Castro-Tirado}, {Cwiek}, {Claret}, {Opiela},
  {{\.Z}arnecki}, {Gorosabel}, {Oates}, {Cunniffe}, {Jeong}, {Hudec},
  {Sokolov}, {Makarov}, {Tello}, {Lara-Gil}, {Kub{\'a}nek}, {Guziy}, {Bai},
  {Fan}, {Wang}, \& {Park}}]{Caballero2015}
{Caballero-Garc{\'{\i}}a} M.~D. {et~al.}, 2015, \mnras, 452, 4195

\bibitem[{{Drake} {et~al}\mbox{.}(2014){Drake}, {Osten}, {Page}, {Kennea},
  {Oates}, {Krimm}, \& {Gehrels}}]{Drake2014}
{Drake} S., {Osten} R., {Page} K.~L., {Kennea} J.~A., {Oates} S.~R., {Krimm}
  H., {Gehrels} N., 2014, The Astronomer's Telegram, 6121

\bibitem[{{Eker} {et~al}\mbox{.}(2008){Eker}, {Ak}, {Bilir}, {Do{\v g}ru},
  {T{\"u}ys{\"u}z}, {Soydugan}, {Bak{\i}{\c s}}, {U{\v g}ra{\c s}}, {Soydugan},
  {Erdem}, \& {Demircan}}]{Eker2008}
{Eker} Z. {et~al.}, 2008, \mnras, 389, 1722

\bibitem[{{Emslie} {et~al}\mbox{.}(2005){Emslie}, {Dennis}, {Holman}, \&
  {Hudson}}]{Emslie2005}
{Emslie} A.~G., {Dennis} B.~R., {Holman} G.~D., {Hudson} H.~S., 2005, Journal
  of Geophysical Research (Space Physics), 110, A11103

\bibitem[{{Emslie} {et~al}\mbox{.}(2012){Emslie}, {Dennis}, {Shih},
  {Chamberlin}, {Mewaldt}, {Moore}, {Share}, {Vourlidas}, \&
  {Welsch}}]{Emslie2012}
{Emslie} A.~G. {et~al.}, 2012, \apj, 759, 71

\bibitem[{{Fender} {et~al}\mbox{.}(2015){Fender}, {Anderson}, {Osten},
  {Staley}, {Rumsey}, {Grainge}, \& {Saunders}}]{Fender2015}
{Fender} R.~P., {Anderson} G.~E., {Osten} R., {Staley} T., {Rumsey} C.,
  {Grainge} K., {Saunders} R.~D.~E., 2015, \mnras, 446, L66

\bibitem[{{Gopalswamy} {et~al}\mbox{.}(2012){Gopalswamy}, {Xie}, {Yashiro},
  {Akiyama}, {M{\"a}kel{\"a}}, \& {Usoskin}}]{Gopalswamy2012}
{Gopalswamy} N., {Xie} H., {Yashiro} S., {Akiyama} S., {M{\"a}kel{\"a}} P.,
  {Usoskin} I.~G., 2012, \ssr, 171, 23

\bibitem[{{Gopalswamy} {et~al}\mbox{.}(2004){Gopalswamy}, {Yashiro}, {Krucker},
  {Stenborg}, \& {Howard}}]{Gopalswamy2004}
{Gopalswamy} N., {Yashiro} S., {Krucker} S., {Stenborg} G., {Howard} R.~A.,
  2004, Journal of Geophysical Research (Space Physics), 109, A12105

\bibitem[{{Hawley} \& {Pettersen}(1991)}]{Hawley1991}
{Hawley} S.~L., {Pettersen} B.~R., 1991, \apj, 378, 725

\bibitem[{{Hinton} \& {Aharonian}(2007)}]{Hinton2007}
{Hinton} J.~A., {Aharonian} F.~A., 2007, \apj, 657, 302

\bibitem[{{Hinton} \& {Hofmann}(2009)}]{Hinton2009}
{Hinton} J.~A., {Hofmann} W., 2009, \araa, 47, 523

\bibitem[{{Isola} {et~al}\mbox{.}(2007){Isola}, {Favata}, {Micela}, \&
  {Hudson}}]{Isola2007}
{Isola} C., {Favata} F., {Micela} G., {Hudson} H.~S., 2007, \aap, 472, 261

\bibitem[{{Kahler}(2001)}]{Kahler2001}
{Kahler} S.~W., 2001, \jgr, 106, 20947

\bibitem[{{Kay} {et~al}\mbox{.}(2016){Kay}, {Opher}, \& {Kornbleuth}}]{Kay2016}
{Kay} C., {Opher} M., {Kornbleuth} M., 2016, \apj, 826, 195

\bibitem[{{Knizhnik} {et~al}\mbox{.}(2011){Knizhnik}, {Swisdak}, \&
  {Drake}}]{Knizhnik2011}
{Knizhnik} K., {Swisdak} M., {Drake} J.~F., 2011, \apjl, 743, L35

\bibitem[{{Kozarev} \& {Schwadron}(2016)}]{Kozarev2016}
{Kozarev} K.~A., {Schwadron} N.~A., 2016, \apj, 831, 120

\bibitem[{{Laming} {et~al}\mbox{.}(2013){Laming}, {Moses}, {Ko}, {Ng},
  {Rakowski}, \& {Tylka}}]{Laming2013}
{Laming} J.~M., {Moses} J.~D., {Ko} Y.-K., {Ng} C.~K., {Rakowski} C.~E.,
  {Tylka} A.~J., 2013, \apj, 770, 73

\bibitem[{{Leitzinger} {et~al}\mbox{.}(2011){Leitzinger}, {Odert}, {Ribas},
  {Hanslmeier}, {Lammer}, {Khodachenko}, {Zaqarashvili}, \&
  {Rucker}}]{Leitzinger2011}
{Leitzinger} M., {Odert} P., {Ribas} I., {Hanslmeier} A., {Lammer} H.,
  {Khodachenko} M.~L., {Zaqarashvili} T.~V., {Rucker} H.~O., 2011, \aap, 536,
  A62

\bibitem[{{Loh} {et~al}\mbox{.}(2017){Loh}, {Corbel}, \& {Dubus}}]{Loh2017}
{Loh} A., {Corbel} S., {Dubus} G., 2017, ArXiv e-prints

\bibitem[{{Mewaldt} {et~al}\mbox{.}(2012){Mewaldt}, {Looper}, {Cohen},
  {Haggerty}, {Labrador}, {Leske}, {Mason}, {Mazur}, \& {von
  Rosenvinge}}]{Mewaldt2012}
{Mewaldt} R.~A. {et~al.}, 2012, \ssr, 171, 97

\bibitem[{{Mirzoyan}(2014)}]{MAGIC2014}
{Mirzoyan} R., 2014, GRB Coordinates Network, 16238

\bibitem[{{Ng} \& {Reames}(2008)}]{Ng2008}
{Ng} C.~K., {Reames} D.~V., 2008, \apjl, 686, L123

\bibitem[{{Osten} {et~al}\mbox{.}(2007){Osten}, {Drake}, {Tueller}, {Cummings},
  {Perri}, {Moretti}, \& {Covino}}]{Osten2007}
{Osten} R.~A., {Drake} S., {Tueller} J., {Cummings} J., {Perri} M., {Moretti}
  A., {Covino} S., 2007, \apj, 654, 1052

\bibitem[{{Osten} {et~al}\mbox{.}(2010){Osten}, {Godet}, {Drake}, {Tueller},
  {Cummings}, {Krimm}, {Pye}, {Pal'shin}, {Golenetskii}, {Reale}, {Oates},
  {Page}, \& {Melandri}}]{Osten2010}
{Osten} R.~A. {et~al.}, 2010, \apj, 721, 785

\bibitem[{{Osten} {et~al}\mbox{.}(2016){Osten}, {Kowalski}, {Drake}, {Krimm},
  {Page}, {Gazeas}, {Kennea}, {Oates}, {Page}, {de Miguel}, {Nov{\'a}k},
  {Apeltauer}, \& {Gehrels}}]{Osten2016}
{Osten} R.~A. {et~al.}, 2016, \apj, 832, 174

\bibitem[{{Pesce-Rollins} {et~al}\mbox{.}(2015){Pesce-Rollins}, {Omodei},
  {Petrosian}, {Liu}, {Rubio da Costa}, {Allafort}, \&
  {Chen}}]{Pesce-Rollins2015}
{Pesce-Rollins} M., {Omodei} N., {Petrosian} V., {Liu} W., {Rubio da Costa} F.,
  {Allafort} A., {Chen} Q., 2015, \apjl, 805, L15

\bibitem[{{Ramaty} {et~al}\mbox{.}(1995){Ramaty}, {Mandzhavidze}, {Kozlovsky},
  \& {Murphy}}]{Ramaty1995}
{Ramaty} R., {Mandzhavidze} N., {Kozlovsky} B., {Murphy} R.~J., 1995, \apjl,
  455, L193

\bibitem[{{Reames}(2000)}]{Reames2000}
{Reames} D.~V., 2000, 26th International Cosmic Ray Conference, ICRC XXVI, 516,
  289

\bibitem[{{Reames}(2013)}]{Reames2013}
{Reames} D.~V., 2013, \ssr, 175, 53

\bibitem[{{Segura} {et~al}\mbox{.}(2010){Segura}, {Walkowicz}, {Meadows},
  {Kasting}, \& {Hawley}}]{Segura2010}
{Segura} A., {Walkowicz} L.~M., {Meadows} V., {Kasting} J., {Hawley} S., 2010,
  Astrobiology, 10, 751

\bibitem[{{Takahashi} {et~al}\mbox{.}(2016){Takahashi}, {Mizuno}, \&
  {Shibata}}]{Takahashi2016}
{Takahashi} T., {Mizuno} Y., {Shibata} K., 2016, \apjl, 833, L8

\bibitem[{{Tsuboi} {et~al}\mbox{.}(2016){Tsuboi}, {Yamazaki}, {Sugawara},
  {Kawagoe}, {Kaneto}, {Iizuka}, {Matsumura}, {Nakahira}, {Higa}, {Matsuoka},
  {Sugizaki}, {Ueda}, {Kawai}, {Morii}, {Serino}, {Mihara}, {Tomida}, {Ueno},
  {Negoro}, {Daikyuji}, {Ebisawa}, {Eguchi}, {Hiroi}, {Ishikawa}, {Isobe},
  {Kawasaki}, {Kimura}, {Kitayama}, {Kohama}, {Kotani}, {Nakagawa}, {Nakajima},
  {Ozawa}, {Shidatsu}, {Sootome}, {Sugimori}, {Suwa}, {Tsunemi}, {Usui},
  {Yamamoto}, {Yamaoka}, \& {Yoshida}}]{Tsuboi2016}
{Tsuboi} Y. {et~al.}, 2016, \pasj, 68, 90

\bibitem[{{Venot} {et~al}\mbox{.}(2016){Venot}, {Rocchetto}, {Carl}, {Roshni
  Hashim}, \& {Decin}}]{Venot2016}
{Venot} O., {Rocchetto} M., {Carl} S., {Roshni Hashim} A., {Decin} L., 2016,
  \apj, 830, 77

\bibitem[{{Zhou} {et~al}\mbox{.}(2017){Zhou}, {Ng}, {Beacom}, \&
  {Peter}}]{Zhou2017}
{Zhou} B., {Ng} K.~C.~Y., {Beacom} J.~F., {Peter} A.~H.~G., 2017, \prd, 96,
  023015

\end{thebibliography}
\label{lastpage}

\end{document}